\documentclass[aps,prb,twocolumn,superscriptaddress,showpacs]{revtex4}
\usepackage{graphicx}
\usepackage{graphicx,color}
\usepackage{epstopdf}
\usepackage{amsfonts}
\usepackage{amsmath,amssymb,mathrsfs,amsbsy}
\usepackage{bm}
\usepackage{float}
\usepackage{pst-grad}
\usepackage{dcolumn}
\usepackage{units}
\usepackage{nicefrac}
\usepackage{array}
\usepackage{booktabs}
\usepackage{hhline}
\newcommand{\Lim}[1]{\raisebox{0.5ex}{\scalebox{0.8}{$\displaystyle \lim_{#1}\;$}}}
\newcommand\Tstrut{\rule{0pt}{2.5ex}}
\newcommand\Bstrut{\rule[-4ex]{0pt}{0pt}}

\begin{document}

\title{Symmetry constraints on the elastoresistivity tensor}

\author{M. C. Shapiro}
\email[Corresponding author: ]{maxshaps@stanford.edu}
\affiliation{Stanford Institute for Materials and Energy Sciences, SLAC National Accelerator Laboratory,\\ 2575 Sand Hill Road, Menlo Park, California 94025, USA} 
\affiliation{Geballe Laboratory for Advanced Materials and Department of Applied Physics, Stanford University, Stanford, California 94305, USA}
\author{Patrik Hlobil}
\affiliation{Karlsruher Institut f\"{u}r Technologie, Institut f\"{u}r Theorie der Kondensierten Materie, 76128 Karlsruhe, Germany}
\author{A. T. Hristov}
\affiliation{Stanford Institute for Materials and Energy Sciences, SLAC National Accelerator Laboratory,\\ 2575 Sand Hill Road, Menlo Park, California 94025, USA} 
\affiliation{Geballe Laboratory for Advanced Materials and Department of Physics, Stanford University, Stanford, California 94305, USA}
\author{Akash V. Maharaj}
\affiliation{Stanford Institute for Materials and Energy Sciences, SLAC National Accelerator Laboratory,\\ 2575 Sand Hill Road, Menlo Park, California 94025, USA} 
\affiliation{Geballe Laboratory for Advanced Materials and Department of Physics, Stanford University, Stanford, California 94305, USA} 
\author{I. R. Fisher}
\affiliation{Stanford Institute for Materials and Energy Sciences, SLAC National Accelerator Laboratory,\\ 2575 Sand Hill Road, Menlo Park, California 94025, USA} 
\affiliation{Geballe Laboratory for Advanced Materials and Department of Applied Physics, Stanford University, Stanford, California 94305, USA}

\begin{abstract}

The elastoresistivity tensor $m_{ij,kl}$ characterizes changes in a material's resistivity due to strain.  As a fourth-rank tensor, elastoresistivity can be a uniquely useful probe of the symmetries and character of the electronic state of a solid.  We present a symmetry analysis of $m_{ij,kl}$ (both in the presence and absence of a magnetic field) based on the crystalline point group, focusing for pedagogic purposes on the $D_{4h}$ point group (of relevance to several materials of current interest).  We also discuss the relation between $m_{ij,kl}$ and various thermodynamic susceptibilities, particularly where they are sensitive to critical fluctuations proximate to a critical point at which a point group symmetry is spontaneously broken.

\end{abstract}
\pacs{72.15.-v,72.20.Fr,75.40.-s,75.47.-m}
\maketitle

\section{Introduction}

Although the occurrence of a phase transition in a material is often reflected by anomalies in its resistivity, such transport measurements do not generally identify the precise nature of the underlying broken symmetry. However, as a derivative of the resistivity, the \textit{elastoresistivity}---a fourth-rank tensor that linearly relates normalized resistivity changes and strain---is also sensitive to directional anisotropies and other point group symmetries which more subtly manifest in the resistivity itself. Yet despite its importance in the semiconductor industry,\cite{sun_2010} elastoresistivity has only recently been exploited as a probe of broken symmetry in the field of strongly correlated electron systems.\cite{chu_2012,kuo_2013,kuo_2014,riggs_2015,watson_2015}  Since the electrons in these materials are often strongly coupled to the crystal lattice as compared to simple metals, and because transport measurements are sensitive to long wavelength electronic excitations at the Fermi level, elastoresistivity is a potentially valuable tool in elucidating the nature of broken symmetries in these complex systems.\cite{hlobil_2015}

By measuring the temperature dependence of the in-plane elastoresistance, recent experiments have probed the nematic susceptibility of a series of iron-pnictide\cite{chu_2012,kuo_2013,kuo_2014} and heavy fermion\cite{riggs_2015} superconductors, signaling in both cases the nematic character of the fluctuations associated with the underlying order parameter.  However, earlier discussions were limited in scope, considering the zero magnetic field limit and mostly reasoning by analogy with the elastic stiffness tensor.  To advance the technique, what is needed is a full theoretical exposition of the structure of the elastoresistivity tensor, including its symmetry constraints and its magnetic field dependence.

In this manuscript, we pedagogically discuss the constraints that symmetry imposes on the elastoresistivity tensor, both in the presence and absence of an externally applied magnetic field.  Several aspects of this treatment are nontrivial due to the different ways in which the resistivity and strain tensors transform.  To our knowledge, this is the first discussion of the symmetry properties of the full elastoresistivity tensor in the presence of a magnetic field or for a point group other than cubic $O_h$.  We illustrate this with the specific point group of $D_{4h}$ but emphasize that the symmetry principles that are outlined can be straightforwardly generalized to any point group.  Given the constraints imposed by directly inherited and point group symmetries (Section \ref{sec:definitionandsymmetries}), we derive in Section \ref{sec:d4htensor} the explicit form of the elastoresistivity tensor in $D_{4h}$ (Eq. \eqref{eq:13}).  We then discuss how particular combinations of elastoresistivity coefficients are related to various thermodynamic susceptibilities of the material within the framework of the Landau paradigm of phase transitions (Section \ref{sec:susceptibilities}); it is this connection that makes elastoresistivity a powerful experimental quantity for determining the symmetry of an order parameter for a continuous phase transition.

\section{Definition and Inherited Symmetries}\label{sec:definitionandsymmetries}

The elastoresistivity tensor $m_{ij,kl}(\boldsymbol H )$ is of fourth-rank and linearly relates the (normalized) strain-induced resistivity change $\left(\nicefrac{\Delta \rho}{\rho} \right)_{ij}(\boldsymbol H)$ and the strain $\epsilon_{kl}$ according to

\begin{equation}
\label{eq:1}
m_{ij,kl}(\boldsymbol H ) \equiv \frac{\partial \left(\nicefrac{\Delta \rho}{\rho}\right)_{ij}(\boldsymbol H)}{\partial \epsilon_{kl}} \bigg|_{\hat{\epsilon} = \hat{0}},
\end{equation}

\noindent where we write $\left(\nicefrac{\Delta \rho}{\rho}\right)_{ij}(\boldsymbol H)$ and $\epsilon_{kl}$ as the nine component vectors

\vspace{\baselineskip}
$\left(\nicefrac{\Delta \rho}{\rho}\right)_{ij}(\boldsymbol H) = \begin{pmatrix} \left(\nicefrac{\Delta \rho}{\rho}\right)_{xx}(\boldsymbol H) \\ \left(\nicefrac{\Delta \rho}{\rho}\right)_{yy}(\boldsymbol H) \\ \left(\nicefrac{\Delta \rho}{\rho}\right)_{zz}(\boldsymbol H) \\ \left(\nicefrac{\Delta \rho}{\rho}\right)_{yz}(\boldsymbol H) \\ \left(\nicefrac{\Delta \rho}{\rho}\right)_{zy}(\boldsymbol H) \\ \left(\nicefrac{\Delta \rho}{\rho}\right)_{zx}(\boldsymbol H) \\ \left(\nicefrac{\Delta \rho}{\rho}\right)_{xz}(\boldsymbol H) \\ \left(\nicefrac{\Delta \rho}{\rho}\right)_{xy}(\boldsymbol H) \\ \left(\nicefrac{\Delta \rho}{\rho}\right)_{yx}(\boldsymbol H) \end{pmatrix} \hspace{2mm} $and$ \hspace{2mm} \epsilon_{kl} = \begin{pmatrix} \epsilon_{xx} \\ \epsilon_{yy} \\ \epsilon_{zz} \\ \epsilon_{yz} \\ \epsilon_{zy} \\ \epsilon_{zx} \\ \epsilon_{xz} \\ \epsilon_{xy} \\ \epsilon_{yx} \end{pmatrix}$
\vspace{\baselineskip}

\noindent in order to represent $m_{ij,kl}(\boldsymbol H )$ as a $9 \times 9$ matrix.  Whereas the strain tensor is defined in a manifestly symmetric manner\cite{footnote1} ($\epsilon_{kl} \equiv \frac{1}{2}(\frac{\partial u_k}{\partial x_l}+\frac{\partial u_l}{\partial x_k})$) and so there is no need to distinguish between off-diagonal terms (e.g., $\epsilon_{zx} = \epsilon_{xz}$), the same is not generally true of changes in resistivity in a magnetic field, where for example the Hall effect explicitly requires $(\nicefrac{\Delta \rho}{\rho})_{ij}(\boldsymbol H) \neq (\nicefrac{\Delta \rho}{\rho})_{ji}(\boldsymbol H)$ for finite $\boldsymbol H$; therefore, we choose not to use the compactified Voigt notation and instead include all nine components of both the change in resistivity and strain tensors.

As written, there is some ambiguity about the normalization constant $\rho$ in each component of the change in resistance tensor, and in particular in the off-diagonal terms where $\rho_{ij} = 0$ in vanishing magnetic field.  From the perspective of symmetry, and in order to preserve the transformation properties of $\left(\nicefrac{\Delta \rho}{\rho}\right)_{ij}(\boldsymbol H)$ as a second-rank tensor, the following normalization scheme is motivated.  Perturbatively, the strained resistivity tensor $\rho_{ij}(\hat{\epsilon})$ is equal to the unstrained resistivity $\rho_{ij}(\hat{\epsilon}=\hat{0})$ plus a strain-induced resistivity change $\Delta\rho_{ij}(\hat{\epsilon})$:

\begin{equation}
\label{eq:2}
\rho_{ij}(\hat{\epsilon}) = \rho_{ij}(\hat{\epsilon}=\hat{0}) + \Delta\rho_{ij}(\hat{\epsilon}).
\end{equation}

\noindent If the tensors were scalars, we would unambiguously factor out $\rho$ to define $\nicefrac{\Delta\rho}{\rho}$; however, since $\rho_{ij}(\hat{\epsilon}=\hat{0})$ is a tensor which does not generally commute with $\Delta\rho_{ij}(\hat{\epsilon})$, one would obtain a different result depending on whether $\rho_{ij}(\hat{\epsilon}=\hat{0})$ was factored on the left or right.  Instead, we define the symmetric factorization

\begin{equation}
\label{eq:3}
\left( \nicefrac{\boldsymbol{\Delta \rho}}{\boldsymbol{\rho}} \right) \equiv (\boldsymbol{\rho}(\hat{\epsilon}=\hat{0}))^{-1/2} \cdot \boldsymbol{{\Delta\rho}}(\hat{\epsilon}) \cdot (\boldsymbol{\rho}(\hat{\epsilon}=\hat{0}))^{-1/2},
\end{equation}

\noindent which in the limit of $\rho_{ii} \gg \rho_{ij}$  $(i \neq j)$ results in

\begin{equation}
\label{eq:4}
(\nicefrac{\Delta \rho}{\rho})_{ij} \equiv (\nicefrac{\Delta \rho_{ij}}{\sqrt{\rho_{ii}}{\sqrt{\rho_{jj}}}}).
\end{equation}

\noindent This normalization scheme preserves the transformation properties of $(\nicefrac{\Delta \rho}{\rho})_{ij}$ as a second-rank tensor and ensures that all symmetries of $\Delta\rho_{ij}(\hat{\epsilon})$ are retained in $(\nicefrac{\Delta \rho}{\rho})_{ij}$ (i.e., preserves the group structure of $\Delta\rho_{ij}(\hat{\epsilon})$).

\subsection{Directly Inherited Symmetries}

The symmetry properties possessed by the change in resistivity and strain tensors are retained by the elastoresistivity tensor as well, and correspondingly constrain the number of independent coefficients.  For the change in resistivity tensor, the relevant symmetry is given by the Onsager relationship\cite{onsager_1931,footnote2} $(\nicefrac{\Delta \rho}{\rho})_{ij}(\boldsymbol H) = (\nicefrac{\Delta \rho}{\rho})_{ji}(-\boldsymbol H)$, which directly implies that $m_{ij,kl}(\boldsymbol H ) = m_{ji,kl}(-\boldsymbol H )$.  The symmetry of the strain tensor $\epsilon_{kl} = \epsilon_{lk}$ transfers as well, requiring that $m_{ij,kl}(\boldsymbol H ) = m_{ij,lk}(\boldsymbol H )$.  These symmetries reduce the 81 independent $m_{ij,kl}(\boldsymbol H )$ to 54.

\subsection{Point Group Symmetry Constraints}

In addition to the directly inherited symmetries, the form of the elastoresistivity tensor depends on the point group symmetry of the crystal lattice and the presence and direction of a magnetic field. Under generic coordinate transformations, the elastoresistivity tensor is transformed as $m \rightarrow m^{\prime}$ according to

\begin{align}
\label{eq:5}
m^{\prime}_{ij,kl} = O_{ia} O_{jb} O_{kc} O_{ld}\, m_{ab,cd}\,\,,
\end{align}

\noindent where $O_{ia}$ is the appropriate transformation matrix relating the two coordinate systems.\cite{footnote3} However, when the coordinate transformation is a group element of the crystalline point group, this physical response function is necessarily invariant, i.e., $m^{\prime}_{ij,kl} = m_{ij,kl}$, and this equation leads to constraints on the individual elements of the elastoresistivity tensor. This is the essence of Neumann's principle.

From hereon, we will specialize to symmetry operations of the point group $D_{4h}$, which consists of 16 symmetry elements involving mirrors, rotations, and improper rotations.  Our motivation for choosing this particular point group to illustrate the symmetry properties of the elastoresistivity tensor is due to the fact that several materials of current interest have such a symmetry; in particular, but not exclusively, materials that have the common ThCr$_2$Si$_2$ structure type crystallize in this point group.  However, the symmetry considerations that we outline below can be readily applied to other point groups.  Generally, the presence of a finite $\boldsymbol H$ reduces the symmetry of the point group to a subgroup of $D_{4h}$, with the particular subgroup depending on the orientation of the field relative to the primitive axes;\cite{footnote4} for simplicity and throughout, we will assume that all magnetic fields are oriented along the primary rotation axis (i.e., $c$ axis) of the crystal, which we choose to label as the $z$ direction (i.e., $\boldsymbol H = H_z \hat{z}$).  To derive the symmetry constraints imposed by the point group, we need only consider the set of all rotations and reflections that are group elements;\cite{footnote5} we now consider all such elements of $D_{4h}$ independently and enumerate their implications for the elastoresistivity tensor.

First, consider the mirror operation about the $xy$ plane (denoted by $\sigma_{z}$), which is represented in matrix form (with a Cartesian basis) as

\begin{equation}
\label{eq:6}
\sigma_{z} = \begin{pmatrix} 1 & 0 & 0 \\ 0 & 1 & 0 \\ 0 & 0 & -1 \end{pmatrix}.
\end{equation}

\noindent $\sigma_z$ takes $z \to -z$ while leaving the other spatial dimensions unchanged and correspondingly transforms the elastoresistivity tensor according to $m_{ij,kl}(H_z \hat{z}) \xrightarrow{\sigma_{z}} (-1)^{\mathcal{N}_{z}} m_{ij,kl}(\sigma_{z} H_z \hat{z} )$, where $\mathcal{N}_{z}$ is the number of times a $z$ index appears among the $\{i,j,k,l\}$.  In order to discern how the mirror operation $\sigma_{\hat{n}}$ acts on the magnetic field (where $\hat{n}$ denotes the normal vector to the mirror plane), we decompose its action into an inversion and a rotation by $\pi$ radians about the $\hat{n}$ axis: $\sigma_{\hat{n}} = O_{\hat{n}}(\pi) \cdot \mathcal{I}$.  Since the magnetic field is a pseudovector, it rotates as a regular vector but is invariant under inversion; therefore, since $O_{\hat{z}}(\pi) H_z \hat{z} = H_z \hat{z}$, $\sigma_z H_z \hat{z} = H_z \hat{z}$ and for a crystal with mirror symmetry about the $xy$ plane

\begin{equation}
\label{eq:7}
m_{ij,kl}(H_z) = (-1)^{\mathcal{N}_{z}} m_{ij,kl}(H_z) \quad \Big[ \sigma_{z} \Big].
\end{equation}

There are analogous relations for the mirror symmetries about the $yz$ and $xz$ planes as well ($\sigma_{x}$ and $\sigma_{y}$, respectively, where again the subscripts denote normal directions to the mirror plane), although care must be taken to account for the transformation of the magnetic field.  Since $\sigma_{\hat{x}} H_{z} \hat{z}  = \sigma_{\hat{y}} H_z \hat{z} = -H_z \hat{z}$, these mirror symmetries then require

\begin{align}
\label{eq:8}
&m_{ij,kl}(H_z) = (-1)^{\mathcal{N}_x} m_{ij,kl}(-H_z) \quad \Big[ \sigma_{x} \Big] \\
&m_{ij,kl}(H_z) = (-1)^{\mathcal{N}_y} m_{ij,kl}(-H_z) \quad \Big[ \sigma_{y} \Big]. \nonumber
\end{align}

The final mirror symmetries contained in $D_{4h}$ are the diagonal reflections about the planes spanned by the lines $x=\pm y$ and the $z$ axis.  These symmetry operations (denoted $\sigma_{x=y}$ and $\sigma_{x=-y}$, respectively, where again the subscripts denote normal directions to the mirror plane) take (in the Cartesian basis) the matrix form

\begin{equation}
\label{eq:9}
\sigma_{x=y} = \begin{pmatrix} 0 & -1 & 0 \\ -1 & 0 & 0 \\ 0 & 0 & 1 \end{pmatrix}, \quad \sigma_{x=-y} = \begin{pmatrix} 0 & 1 & 0 \\ 1 & 0 & 0 \\ 0 & 0 & 1 \end{pmatrix}
\end{equation}

\noindent and transform the magnetic field as $\sigma_{x = \pm y} H_z \hat{z} = -H_z \hat{z}$; correspondingly, the elastoresistivity tensor is constrained according to

\begin{align}
\label{eq:10}
&m_{ij,kl}(H_z) = (-1)^{\mathcal{N}_x + \mathcal{N}_y} m_{ij,kl}(-H_z) \bigg|_{x \leftrightarrow y} \quad \Big[ \sigma_{x=y} \Big] \nonumber \\
&m_{ij,kl}(H_z) = m_{ij,kl}(-H_z) \bigg|_{x \leftrightarrow y} \quad \Big[ \sigma_{x=-y} \Big],
\end{align}

\noindent where $\big|_{x \leftrightarrow y}$ conveys that all initial $x$ and $y$ indices are to be interchanged upon the symmetry operation.

The final set of symmetry constraints is imposed by rotational symmetries.  $D_{4h}$ possesses a primary fourfold rotational symmetry about the $z$ axis ($C_4$), two secondary twofold rotations about the $x$ and $y$ axes ($C_2'(x)$ and $C_2'(y)$, respectively), and two tertiary twofold rotations about the lines $x=\pm y$ ($C_2''(1)$ and $C_2''(2)$, respectively).  The operations are represented in matrix form (in the Cartesian basis) as

\begin{align}
\label{eq:11}
& C_4 = \begin{pmatrix} 0 & -1 & 0 \\ 1 & 0 & 0 \\ 0 & 0 & 1 \end{pmatrix}, & C_2'(x) = \begin{pmatrix} 1 & 0 & 0 \\ 0 & -1 & 0 \\ 0 & 0 & -1 \end{pmatrix}, \nonumber \\
& C_2'(y) = \begin{pmatrix} -1 & 0 & 0 \\ 0 & 1 & 0 \\ 0 & 0 & -1 \end{pmatrix}, & C_2''(1) = \begin{pmatrix} 0 & 1 & 0 \\ 1 & 0 & 0 \\ 0 & 0 & -1 \end{pmatrix}, \nonumber \\
& C_2''(2) = \begin{pmatrix} 0 & -1 & 0 \\ -1 & 0 & 0 \\ 0 & 0 & -1 \end{pmatrix}. &  
\end{align}

\noindent The primary rotational symmetry preserves the magnetic field orientation while the secondary and tertiary rotations invert the field, and so the rotational symmetry operations collectively require that the elastoresistivity tensor obey

\begin{align}
\label{eq:12}
m_{ij,kl}(H_z) &= (-1)^{\mathcal{N}_{y}} m_{ij,kl}(H_z) \bigg|_{x \leftrightarrow y} \qquad \qquad \hspace{1mm} \Big[C_4 \Big] \nonumber \\
m_{ij,kl}(H_z) &= (-1)^{\mathcal{N}_{y}+\mathcal{N}_{z}} m_{ij,kl}(-H_z) \qquad \qquad \hspace{-0.5mm} \Big[ C_2'(x) \Big] \nonumber \\
m_{ij,kl}(H_z) &= (-1)^{\mathcal{N}_{x}+\mathcal{N}_{z}} m_{ij,kl}(-H_z) \quad \Big[ C_2'(y) \Big] \\
m_{ij,kl}(H_z) &= (-1)^{\mathcal{N}_{z}} m_{ij,kl}(-H_z) \bigg|_{x \leftrightarrow y} \qquad \quad \hspace{1.5mm} \Big[ C_2''(1) \Big] \nonumber \\
m_{ij,kl}(H_z) &= (-1)^{\mathcal{N}_{x}+\mathcal{N}_{y}+\mathcal{N}_{z}} m_{ij,kl}(-H_z) \bigg|_{x \leftrightarrow y} \Big[ C_2''(2) \Big]. \nonumber
\end{align}

\begin{table*}
\caption{\textbf{Elastoresistivity Symmetry Properties}}
\centering
(Assuming $\boldsymbol H = H_z \hat{z}$)
\makebox[\textwidth][c]{
\begin{tabular}{c | c}
\hline \hline
Principle / Symmetry & Elastoresistivity Constraint \Tstrut \\[1ex]
\hline \hline
Onsager & $m_{ij,kl}(H_z) = m_{ji,kl}(-H_z)$ \Tstrut \\
\hline
Strain Definition & $m_{ij,kl}(H_z) = m_{ij,lk}(H_z)$ \Tstrut \\
\hline
$\sigma_x$ Mirror & $m_{ij,kl}(H_z) = (-1)^{\mathcal{N}_{x}} m_{ij,kl}(-H_z)$ \Tstrut \\
\hline
$\sigma_y$ Mirror & $m_{ij,kl}(H_z) = (-1)^{\mathcal{N}_{y}} m_{ij,kl}(-H_z)$ \Tstrut \\
\hline
$\sigma_z$ Mirror & $m_{ij,kl}(H_z) = (-1)^{\mathcal{N}_{z}} m_{ij,kl}(H_z)$ \Tstrut \\
\hline
$\sigma_{x=y}$ Mirror & $m_{ij,kl}(H_z) = (-1)^{\mathcal{N}_x + \mathcal{N}_y} m_{ij,kl}(-H_z) \bigg|_{x \leftrightarrow y}$ \Tstrut \\
\hline
$\sigma_{x=-y}$ Mirror & $m_{ij,kl}(H_z) = m_{ij,kl}(-H_z) \bigg|_{x \leftrightarrow y}$ \Tstrut \\
\hline
$C_4$ Rotation & $m_{ij,kl}(H_z) = (-1)^{\mathcal{N}_{y}} m_{ij,kl}(H_z) \bigg|_{x \leftrightarrow y}$ \Tstrut \\
\hline
$C_2'(x)$ Rotation & $m_{ij,kl}(H_z) = (-1)^{\mathcal{N}_{y}+\mathcal{N}_{z}} m_{ij,kl}(-H_z)$ \Tstrut \\
\hline
$C_2'(y)$ Rotation & $m_{ij,kl}(H_z) = (-1)^{\mathcal{N}_{x}+\mathcal{N}_{z}} m_{ij,kl}(-H_z)$ \Tstrut \\
\hline
$C_2''(1)$ Rotation & $m_{ij,kl}(H_z) = (-1)^{\mathcal{N}_{z}} m_{ij,kl}(-H_z) \bigg|_{x \leftrightarrow y}$ \Tstrut \\
\hline
$C_2''(2)$ Rotation & $m_{ij,kl}(H_z) = (-1)^{\mathcal{N}_{x}+\mathcal{N}_{y}+\mathcal{N}_{z}} m_{ij,kl}(-H_z) \bigg|_{x \leftrightarrow y}$ \Tstrut \\
\hline
\end{tabular}}
\label{table1}
\end{table*}

\noindent The totality of these symmetries and their consequences for the elastoresistivity tensor are summed up in Table~\ref{table1}.

\section{Elastoresistivity Tensor for $D_{4h}$}\label{sec:d4htensor}

We now explicitly write down the elastoresistivity tensor for the $D_{4h}$ point group, which possesses all of the symmetries in Table \ref{table1}.  These conditions require that certain coefficients vanish (e.g., $\sigma_z$ symmetry requires $m_{xx,yz}(H_z) = -m_{xx,yz}(H_z) = 0$) while equating certain others (e.g., $C_4$ symmetry requires $m_{yy,xx}(H_z) = m_{xx,yy}(H_z)$).  Imposing all symmetry constraints, the elastoresistivity tensor is given as\cite{footnote6}

\begin{widetext}
\begin{equation}
\label{eq:13}
m^{D_{4h}}_{ij,kl}(H_z) =
\begingroup
\renewcommand*{\arraystretch}{1.25}
\begin{pmatrix}
m_{xx,xx} & m_{xx,yy} & m_{xx,zz} & 0 & 0 & 0 & 0 & 0 & 0 \\
m_{xx,yy} & m_{xx,xx} & m_{xx,zz} & 0 & 0 & 0 & 0 & 0 & 0 \\
m_{zz,xx} & m_{zz,xx} & m_{zz,zz} & 0 & 0 & 0 & 0 & 0 & 0 \\
\hhline{~~~~~--~~}
0 & 0 & 0 & m_{yz,yz} & m_{yz,yz} & \multicolumn{1}{|c}{m_{yz,zx}} & \multicolumn{1}{c|}{m_{yz,zx}} & 0 & 0 \\
0 & 0 & 0 & m_{yz,yz} & m_{yz,yz} & \multicolumn{1}{|c}{-m_{yz,zx}} & \multicolumn{1}{c|}{-m_{yz,zx}} & 0 & 0 \\
\hhline{~~~----~~}
0 & 0 & 0 & \multicolumn{1}{|c}{m_{yz,zx}} & \multicolumn{1}{c|}{m_{yz,zx}} & m_{yz,yz} & m_{yz,yz} & 0 & 0 \\
0 & 0 & 0 & \multicolumn{1}{|c}{-m_{yz,zx}} & \multicolumn{1}{c|}{-m_{yz,zx}} & m_{yz,yz} & m_{yz,yz} & 0 & 0 \\
\hhline{-----~~~~}
\multicolumn{1}{|c}{m_{xy,xx}} & m_{xy,xx} & \multicolumn{1}{c|}{m_{xy,zz}} & 0 & 0 & 0 & 0 & m_{xy,xy} & m_{xy,xy} \\
\multicolumn{1}{|c}{-m_{xy,xx}} & -m_{xy,xx} & \multicolumn{1}{c|}{-m_{xy,zz}} & 0 & 0 & 0 & 0 & m_{xy,xy} & m_{xy,xy}\\
\hhline{---~~~~~~}
\end{pmatrix}.
\endgroup
\end{equation}
\end{widetext}

\noindent This tensor has 10 independent coefficients, all implicitly dependent on the magnetic field.  Those coefficients (of which there are seven) that have an even number of $x$ and an even number of $y$ indices are correspondingly even functions of the magnetic field due to the $\sigma_x$ and $\sigma_y$ symmetry constraints; conversely, those coefficients (of which there are three, demarcated by surrounding boxes) that have an odd number of $x$ and an odd number of $y$ indices are odd functions of the magnetic field (and consequently vanish in zero field).

The asymmetric appearance of $m^{D_{4h}}_{ij,kl}(H_z)$ in \eqref{eq:13} is fundamentally due to the definition of the elastoresistivity tensor in \eqref{eq:1}, which does not generally admit interchanging the $ij$ and $kl$ indices; this stands in contrast to the elastic stiffness tensor $C_{ij,kl}$, which generally does have the symmetry $ij \leftrightarrow kl$ because of its symmetric definition as the second derivative of the elastic energy density $\mathcal{U}$: $C_{ij,kl} \equiv \nicefrac{\partial^2 \mathcal{U}}{\partial \epsilon_{ij} \partial \epsilon_{kl}}$.  The zeros in the upper right corner are enforced by Onsager and either $\sigma_{x}$ or $\sigma_{y}$ symmetry, but these symmetries only constrain the boxed terms to be odd functions of the magnetic field (i.e., they only vanish in zero field).

\section{Connection to Thermodynamic Susceptibilities}\label{sec:susceptibilities}

A primary motivation for measuring elastoresistivity coefficients is their connection to thermodynamic susceptibilities.  Building on our earlier work\cite{chu_2012,kuo_2013,kuo_2014,riggs_2015,hlobil_2015} and employing a more general formalism, we outline this connection in greater detail.

Continuous phase transitions can be experimentally identified by the observation of a diverging thermodynamic susceptibility across a phase boundary. For example, in the thermal phase transition of an Ising ferromagnet with a Curie temperature $\theta_{_C}$, the magnetic susceptibility $\chi_{_M} \equiv \Lim{H \to 0} \frac{d M}{d H}$   (i.e., the rate that the order parameter $M$ changes in response to its conjugate field $H$, the magnetic field) progressively increases on cooling from $T > \theta_{_C}$ until it diverges at the phase transition. Crucially, the measurement of such a susceptibility is only possible when an available external experimental probe has the same symmetry as (and hence is conjugate to) the order parameter describing the phase transition.  Here, we discuss the classes of order parameters which couple nontrivially (i.e., in a manner which can convey symmetry information about the order parameter) to externally applied strain fields.

\newsavebox{\pmat}
\savebox{\pmat}{$\begin{pmatrix} \epsilon_{xx} & 0 & 0 \\ 0 & \epsilon_{yy} & 0 \\ 0 & 0 & \epsilon_{zz} \end{pmatrix} = \epsilon_{_{A_{1g,1}}} \begin{pmatrix} 1 & 0 & 0 \\ 0 & 1 & 0 \\ 0 & 0 & 0 \end{pmatrix} + \epsilon_{_{A_{1g,2}}} \begin{pmatrix} 0 & 0 & 0 \\ 0 & 0 & 0 \\ 0 & 0 & 1 \end{pmatrix} + \epsilon_{_{B_{1g}}} \begin{pmatrix} 1 & 0 & 0 \\ 0 & -1 & 0 \\ 0 & 0 & 0 \end{pmatrix}$}

\begin{figure*}[t]
\includegraphics[clip=true, width=1.75\columnwidth]{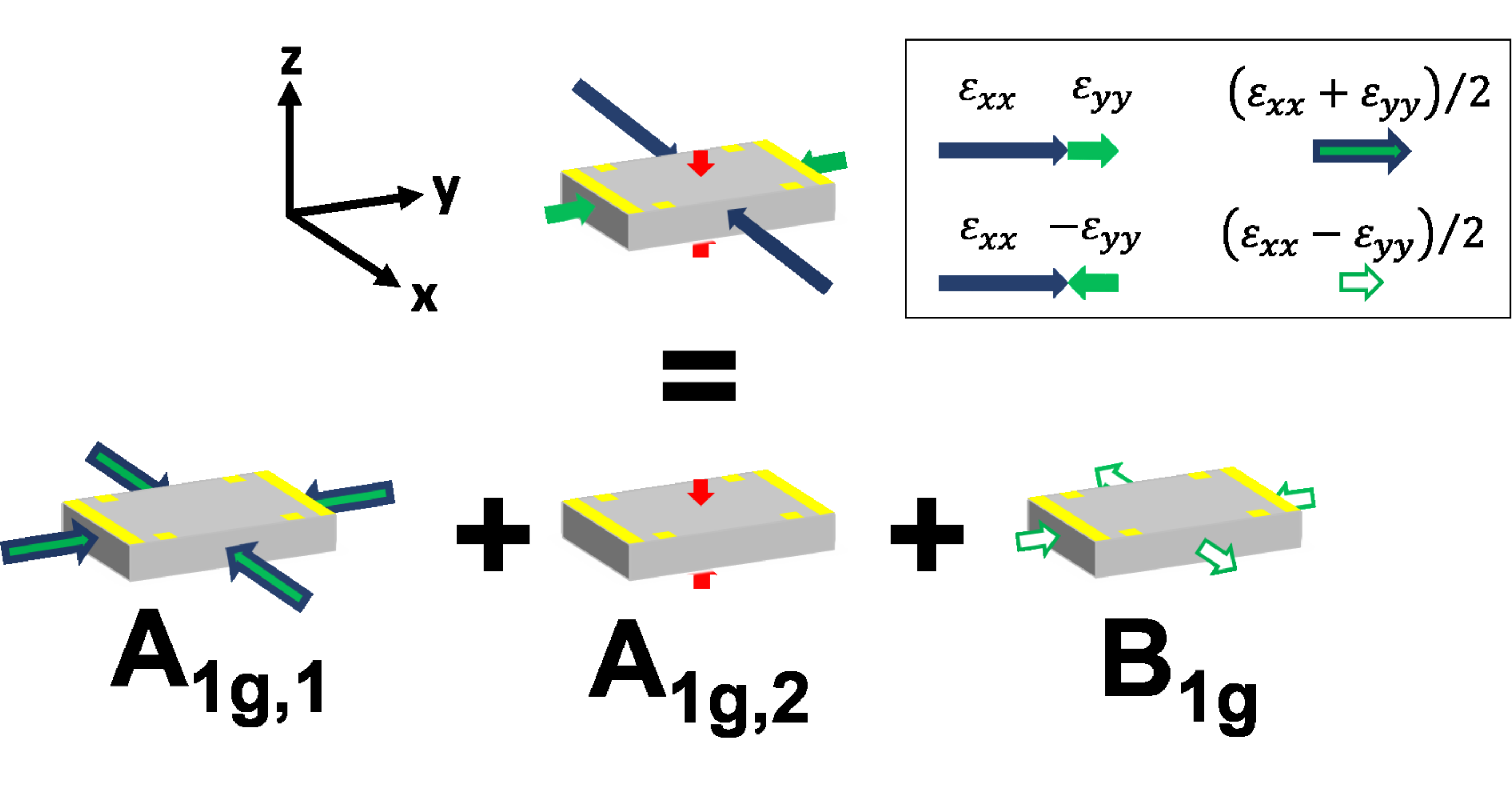}
\caption{Schematic illustration of an arbitrary shearless strain in $D_{4h}$ as decomposed in terms of three irreducible components:\\
\centerline{~\usebox{\pmat},}
\noindent where $\epsilon_{_{A_{1g,1}}} = \frac{1}{2}(\epsilon_{xx} + \epsilon_{yy})$, $\epsilon_{_{A_{1g,2}}} = \epsilon_{zz}$, and $\epsilon_{_{B_{1g}}} = \frac{1}{2}(\epsilon_{xx} - \epsilon_{yy})$. The gray parallelopiped represents a crystalline sample; the yellow regions represent electrical contacts used for transport measurements.}
\label{fig:straindecomposition}
\end{figure*}

\begin{table*}[t]
\caption{\textbf{Irreducible Representations of $D_{4h}$ with Representations in Strain}}
\centering
\makebox[\textwidth][c]{
\begin{tabular}{c | c | c}
\hline\hline
\multicolumn{1}{p{6.6cm}|}{\centering Irreducible Representation\cite{footnote8} \\ (Koster Notation\cite{koster_1957,koster_1963} \& Mulliken Symbol\cite{mulliken_1933,internationalcommission_1955})} & Strain & Strain-Induced Resistivity Change\Tstrut\Bstrut \\
\hline \hline
$\Gamma^+_{1} = A_{1g}$ & $\epsilon_{_{A_{1g,1}}} = \frac{1}{2}(\epsilon_{xx} + \epsilon_{yy})$ & $\left(\nicefrac{\Delta \rho}{\rho}\right)_{_{A_{1g,1}}} \!\! (H_z) = \frac{1}{2} \Big[ \left(\nicefrac{\Delta \rho}{\rho}\right)_{_{xx}}(H_z) + \left(\nicefrac{\Delta \rho}{\rho}\right)_{_{yy}}(H_z) \Big]$ \rule{0pt}{1.5\normalbaselineskip} \\[2ex]
 & $\epsilon_{_{A_{1g,2}}} = \epsilon_{zz}$ & $\left(\nicefrac{\Delta \rho}{\rho}\right)_{_{A_{1g,2}}}(H_z) = \left(\nicefrac{\Delta \rho}{\rho}\right)_{_{zz}}(H_z)$ \rule{0pt}{1.5\normalbaselineskip} \\[2ex]
\hline
$\Gamma^+_{3} = B_{1g}$ & $\epsilon_{_{B_{1g}}} = \frac{1}{2}(\epsilon_{xx} - \epsilon_{yy})$ & $\left(\nicefrac{\Delta \rho}{\rho}\right)_{_{B_{1g}}}(H_z) = \frac{1}{2} \Big[ \left(\nicefrac{\Delta \rho}{\rho}\right)_{_{xx}}(H_z) - \left(\nicefrac{\Delta \rho}{\rho}\right)_{_{yy}}(H_z) \Big]$ \rule{0pt}{1.5\normalbaselineskip} \\[2ex]
\hline
$\Gamma^+_{4} = B_{2g}$ & $\epsilon_{_{B_{2g}}} = \frac{1}{2}(\epsilon_{xy} + \epsilon_{yx}) = \epsilon_{xy}$ & $\left(\nicefrac{\Delta \rho}{\rho}\right)_{_{B_{2g}}}(H_z) = \frac{1}{2} \Big[ \left(\nicefrac{\Delta \rho}{\rho}\right)_{_{xy}}(H_z) + \left(\nicefrac{\Delta \rho}{\rho}\right)_{_{yx}}(H_z) \Big]$ \rule{0pt}{1.5\normalbaselineskip} \\[2ex]
\hline
$\Gamma^+_{5} = E_{g}$ & $\epsilon_{_{E_{g}}} = \dfrac{1}{2} \begin{pmatrix} \epsilon_{xz} + \epsilon_{zx} \\ \epsilon_{yz} + \epsilon_{zy} \end{pmatrix} = \begin{pmatrix} \epsilon_{xz} \\ \epsilon_{yz} \end{pmatrix}$ & $\left(\nicefrac{\Delta \rho}{\rho}\right)_{_{E_{g}}}(H_z) = \dfrac{1}{2} \begin{pmatrix} \left(\nicefrac{\Delta \rho}{\rho}\right)_{_{xz}}(H_z) + \left(\nicefrac{\Delta \rho}{\rho}\right)_{_{zx}}(H_z) \\ \left(\nicefrac{\Delta \rho}{\rho}\right)_{_{yz}}(H_z) + \left(\nicefrac{\Delta \rho}{\rho}\right)_{_{zy}}(H_z) \end{pmatrix}$ \rule{0pt}{1.5\normalbaselineskip} \\[2ex]
\hline
\end{tabular}}
\label{table2}
\end{table*}

Within the Landau paradigm of phase transitions, the singular part of the free energy at a symmetry breaking transition is an analytic function of the order parameter which respects all of the symmetries of the disordered phase of the system. Thus, near a generic phase transition, one can expand the free energy in powers of the order parameter(s) and conjugate field(s), with each term transforming as the identity in the space group of the symmetric phase of the crystal. A generic externally applied strain can only break point group symmetries, and so only order parameters (or products of order parameters, i.e., composite order parameters) which break \textit{exclusively} point group symmetries may couple nontrivially to strain. In this instance, nontrivial coupling refers to terms which are \textit{linear} in the strain $\epsilon_{kl}$; terms that are quadratic in strain, with the form $\epsilon^2|\Delta|^2$, are allowed for any order parameter $\Delta$ (since both $\epsilon^2$ and $|\Delta|^2$ transform individually as the identity) and so generically provide no symmetry information (since they are allowed for any strain or order parameter). 
 
We can therefore identify two classes of order parameters for which external strain proves to be a useful probe: (i) a scalar order parameter $\psi$ that breaks only point group symmetries; (ii) a composite order parameter (derived from a vector order parameter $\vec{\phi}$) that breaks only point group symmetries. In the former case, the general form of the free energy expansion to quartic order near a continuous phase transition is

\begin{align}
\label{eq:14}
f(\psi,\epsilon) &= f_0 + \frac{1}{2} a_0(T-T_c) \psi^2_{_{\Gamma_{i}}} + \frac{1}{4} b \psi^4_{_{\Gamma_{i}}} + \frac{1}{2} c \epsilon^2_{_{\Gamma_{i}}} \nonumber \\
& + \frac{1}{4} d \epsilon^4_{_{\Gamma_{i}}} + \lambda \psi_{_{\Gamma_{i}}} \epsilon_{_{\Gamma_{i}}},
\end{align}

\noindent where $f_0$ is a nonsingular contribution to free energy, the quadratic coefficient changes sign at the transition temperature $T_c$,  $b$ and $d$ are positive to assure $f$ is bounded from below, the coefficient $c$ is the symmetry-dictated combination of elastic moduli, and in the bilinear term $\psi_{_{\Gamma_{i}}}$ and $\epsilon_{_{\Gamma_{i}}}$ both belong to the same nontrivial irreducible representation\cite{footnote7} of the point group.  It is this last term which allows one to measure a thermodynamic susceptibility: any measurement sensitive to the strain $\epsilon_{_{\Gamma_{i}}}$ has contributions from the fluctuations of $\psi_{_{\Gamma_i}}$ which diverge at the phase transition.

For concreteness, let us once more consider the specific case of the $D_{4h}$ point group, where an arbitrary strain can be decomposed into five distinct combinations: $\frac{1}{2} \left( \epsilon_{xx} + \epsilon_{yy} \right)$, $\epsilon_{zz}$, $\frac{1}{2} \left( \epsilon_{xx} - \epsilon_{yy} \right)$, $\epsilon_{xy}$, and $\begin{pmatrix} \epsilon_{xz} \\ \epsilon_{yz} \end{pmatrix}$ (Table~\ref{table2}). Only the last three involve a breaking of point group symmetries, and so the bilinear coupling term in \eqref{eq:14} can correspondingly have three forms depending on the particular irrep of the strain, with an associated thermodynamic susceptibility defined by

\begin{equation}
\label{eq:15}
\chi_{_{\Gamma_i}} \equiv \Lim{\epsilon_{\!_{\, \Gamma_i}} \! \to 0} \frac{d \psi_{_{\Gamma_i}}}{d \epsilon_{_{\Gamma_i}}}.
\end{equation}

\noindent Thus, one can experimentally determine the irreducible representation to which $\psi_{_{\Gamma_i}}$ belongs by applying a strain with a definite $\Gamma_i$ character. Any observed divergence in $\chi_{_{\Gamma_i}}$ signals that $\psi$ belongs to $\Gamma_i$, exactly in analogy to the ferromagnetic case.  A recent example of this corresponds to the case of electronic nematic order in Ba(Fe$_{1-x}$Co$_x$)$_2$As$_2$ for which the tetragonal to orthorhombic transition precedes the subsequent magnetic order.\cite{chu_2012,kuo_2013,kuo_2014}

In the second class of theories, while the order parameters break extra symmetries (e.g., translation, time reversal, gauge invariance, etc.), there is some product of the order parameter fields which breaks solely point group symmetries. Focusing once more on systems with the point group $D_{4h}$, we note that despite the breaking of additional symmetries by the order parameter, terms that are linear in strain but quadratic in order parameters are possible when the order parameter is a (generally complex) vector $\vec{\phi} = (\phi_a, \phi_b)$ that transforms like any of the two dimensional ($E_u$ or $E_g$) representations of $D_{4h}$. In such a scenario, the generic form of the free energy is

\begin{align}
\label{eq:16}
f(\phi,\epsilon) &= \frac{1}{2} a_0(T-T_c) \left(|\phi_{a}|^2 + |\phi_{b}|^2 \right)+ \frac{1}{4} b \left(|\phi_{a}|^2 + |\phi_{b}|^2\right)^2 \nonumber \\
&+ \frac{1}{4}g\left(|\phi_{a}|^2 - |\phi_{b}|^2\right)^2 + \frac{1}{2} c \epsilon^2_{_{\Gamma_{i}}}  + \frac{1}{4} d \epsilon^4_{_{\Gamma_{i}}}\\
& + \lambda\left(|\phi_{a}|^2- |\phi_{b}|^2\right) \epsilon_{_{\Gamma_{i}}}, \nonumber
\end{align}

\noindent where the combination $\left(|\phi_{a}|^2 - |\phi_{b}|^2\right)$ transforms as an irreducible representation $\Gamma_{i}$ of the point group. Examples of such a theory include superconducting states with degenerate $p_x$ and $p_y$ symmetry (where the coupling is to $\epsilon_{_{B_{1g}}}$) or an incommensurate charge density wave with wave vectors oriented along the $[110]$ and $[1\bar{1}0]$ directions (where the coupling is to $\epsilon_{_{B_{2g}}}$).  Because there is no longer a bilinear coupling between strain and the order parameter, a diverging susceptibility is not generically measured.  Instead, any measurement which is sensitive to the symmetry class $\Gamma_i$ will track fluctuations of the composite order parameter $\left(|\phi_{a}|^2 - |\phi_{b}|^2\right)$ close to the transition.  While at temperatures far above the transition one anticipates a Curie-Weiss-like temperature dependence (as long as there is a broad fluctuational regime of $\vec{\phi}$), when $\phi_a$ or $\phi_b$ orders at the transition, these fluctuations are proportional to the square of the fundamental order parameter, and so such a measurement is essentially proportional to the singular contribution to the heat capacity associated with this order parameter; heat capacity-like singularities in the susceptibility may then be observed.\cite{riggs_2015}

With such considerations in mind, we now return to the case of an order parameter which breaks solely point group symmetries and discuss its relation to transport measurements.  While the order parameter $\psi$ is strictly a thermodynamic quantity, it is linearly proportional to all other physical quantities (including non-thermodynamic ones) in the same symmetry class for small values of the order parameter.  In particular, if $\psi$ belongs to the $\Gamma_i$ irrep of the point group, then the strain-induced resistivity change in the same symmetry channel scales as $\left(\nicefrac{\Delta \rho}{\rho}\right)_{_{\Gamma_i}} \sim \psi_{_{\Gamma_i}} + \mathcal{O}(\psi_{_{\Gamma_i}}^3)$.  For the strains accessible in $D_{4h}$, the conjugate resistivity change is given in Table~\ref{table2}, and so the corresponding susceptibilities are

\begin{subequations}
\label{eq:17}
\begin{align}
\begin{split}
\chi_{_{B_{1g}}} &{} \propto \Lim{(\epsilon_{_{xx}} - \epsilon_{_{yy}}) \to 0} \frac{d \Big[ \left(\nicefrac{\Delta \rho}{\rho}\right)_{_{xx}}(H_z) - \left(\nicefrac{\Delta \rho}{\rho}\right)_{_{yy}}(H_z) \Big]}{d \Big[ \epsilon_{_{xx}} - \epsilon_{_{yy}} \Big]} \\
& = m_{xx,xx} - m_{xx,yy}
\end{split} \\
\begin{split}
\chi_{_{B_{2g}}} &{} \propto \Lim{(\epsilon_{_{xy}} + \epsilon_{_{yx}}) \! \to 0} \frac{d \Big[ \left(\nicefrac{\Delta \rho}{\rho}\right)_{_{xy}}(H_z) + \left(\nicefrac{\Delta \rho}{\rho}\right)_{_{yx}}(H_z) \Big]}{d \Big[ \epsilon_{_{xy}} + \epsilon_{_{yx}} \Big]} \\
& = 2m_{xy,xy}
\end{split} \\
\begin{split}
\chi_{_{E_{g}}} \propto &{} \Lim{\begin{pmatrix} \epsilon_{_{xz}} + \epsilon_{_{zx}} \\ \epsilon_{_{yz}} + \epsilon_{_{zy}} \end{pmatrix} \! \to 0} \begin{pmatrix} \nicefrac{d \Big[ \left(\nicefrac{\Delta \rho}{\rho}\right)_{_{xz}}(H_z) + \left(\nicefrac{\Delta \rho}{\rho}\right)_{_{zx}}(H_z) \Big]}{d \Big[ \epsilon_{_{xz}} + \epsilon_{_{zx}} \Big] } \\[6pt] \nicefrac{d \Big[ \left(\nicefrac{\Delta \rho}{\rho}\right)_{_{yz}}(H_z) + \left(\nicefrac{\Delta \rho}{\rho}\right)_{_{zy}}(H_z) \Big]}{d \Big[ \epsilon_{_{yz}} + \epsilon_{_{zy}} \Big] } \end{pmatrix} \\
& = \begin{pmatrix} 2 m_{zx,zx} \\ 2 m_{yz,yz} \end{pmatrix} = \begin{pmatrix} 2 m_{yz,yz} \\ 2 m_{yz,yz} \end{pmatrix}.
\end{split}
\end{align}
\end{subequations}

\noindent The ratios appearing in these susceptibilities correspond to select admixtures of elastoresistivity coefficients, and so by measuring symmetry-motivated combinations of the components of the elastoresistivity tensor, we can infer the behavior of the thermodynamic susceptibilities (up to potentially parameter-dependent coefficients of proportionality) and therefore identify the symmetry class of the order parameter.

We mention in passing that a nematic distortion by definition refers to reduced rotational symmetry, which for in-plane distortions in $D_{4h}$ corresponds to the $B_{1g}$ and $B_{2g}$ irreps; it is for this reason that we have referred to $\chi_{_{B_{1g}}}$ and $\chi_{_{B_{2g}}}$ as nematic susceptibilities and have in previous publications (using the Voigt notation\cite{riggs_2015}) denoted them by $\chi_{_{\mathcal{N}_{[100]}}} = m_{11}-m_{12}$ and $\chi_{_{\mathcal{N}_{[110]}}} = 2m_{66}$, respectively.  In principle, there are also nematic susceptibilities in $D_{4h}$ that correspond to the $E_g$ irrep, but these distortions are out-of-plane and respond to out-of-plane shears (see $\epsilon_{_{E_g}}$ in Table \ref{table2}).

\section{Measurements Of Additional Coefficients}

There are two additional classes of elastoresistivity coefficients which, while they do not correspond to thermodynamic susceptibilities, may nevertheless be sensible to measure.  In the first class, time-reversal odd resistive responses to strain (e.g., $\nicefrac{[ \left(\nicefrac{\Delta \rho}{\rho}\right)_{_{yz}}(H_z) - \left(\nicefrac{\Delta \rho}{\rho}\right)_{_{zy}}(H_z) ]}{[ \epsilon_{xz} + \epsilon_{zx} ]}$) cannot probe susceptibilities because time-reversal odd order parameters cannot bilinearly couple to strain; however, such a ratio does correspond to a distinct elastoresistivity coefficient (in this instance, $2 m_{yz,zx}$).  Similarly, in the second class, resistive responses to $A_{1g}$ strains (e.g., $\nicefrac{\left(\nicefrac{\Delta \rho}{\rho}\right)_{_{zz}}}{\epsilon_{zz}}$) also do not correspond to susceptibilities (because $\epsilon_{_{A_{1g}}}$ does not break a symmetry) but can still be related to elastoresistivity coefficients; in this instance, care must be taken because the two $A_{1g}$ strains can cause distinct elastoresistivity coefficients to mix into each other if one of the $A_{1g}$ strains is not constrained to vanish.  For completeness, we have enumerated below additional coefficients in both of these classes:

\begin{subequations}
\label{eq:18}
\begin{align}
\begin{split}
&{} \Lim{(\epsilon_{_{xx}} + \epsilon_{_{yy}}) \to 0} \frac{d \Big[ \left(\nicefrac{\Delta \rho}{\rho}\right)_{_{xx}}(H_z) + \left(\nicefrac{\Delta \rho}{\rho}\right)_{_{yy}}(H_z) \Big]}{d \Big[ \epsilon_{_{xx}} + \epsilon_{_{yy}} \Big]} \bigg|_{\epsilon_{zz}=0} \\
& \quad = m_{xx,xx} + m_{xx,yy}
\end{split} \\
\begin{split}
&{} \Lim{\epsilon_{_{zz}} \! \to 0} \frac{d \Big[ \left(\nicefrac{\Delta \rho}{\rho}\right)_{_{zz}}(H_z) \Big]}{d \Big[ \epsilon_{_{zz}} \Big]} \bigg|_{\epsilon_{xx}+\epsilon_{yy}=0} = m_{zz,zz}
\end{split}
\end{align}
\begin{align}
\begin{split}
&{} \Lim{(\epsilon_{_{xx}} + \epsilon_{_{yy}}) \to 0} \frac{d \Big[ \left(\nicefrac{\Delta \rho}{\rho}\right)_{_{zz}}(H_z) \Big]}{d \Big[ \epsilon_{_{xx}} + \epsilon_{_{yy}} \Big]} \bigg|_{\epsilon_{zz}=0} = m_{zz,xx}
\end{split} \\
\begin{split}
&{} \Lim{\epsilon_{_{zz}} \to 0} \frac{d \Big[ \left(\nicefrac{\Delta \rho}{\rho}\right)_{_{xx}}(H_z) + \left(\nicefrac{\Delta \rho}{\rho}\right)_{_{yy}}(H_z) \Big]}{d \Big[ \epsilon_{zz} \Big]} \bigg|_{\epsilon_{_{xx}} + \epsilon_{_{yy}}=0} \\
& \quad = 2 m_{xx,zz}
\end{split}\\
\begin{split}
&{} \Lim{(\epsilon_{_{xx}} + \epsilon_{_{yy}}) \to 0} \frac{d \Big[ \left(\nicefrac{\Delta \rho}{\rho}\right)_{_{xy}}(H_z) - \left(\nicefrac{\Delta \rho}{\rho}\right)_{_{yx}}(H_z) \Big]}{d \Big[ \epsilon_{_{xx}} + \epsilon_{_{yy}} \Big]} \bigg|_{\epsilon_{zz}=0} \\
& \quad = 2 m_{xy,xx}
\end{split} \\
\begin{split}
&{} \Lim{\epsilon_{_{zz}} \to 0} \frac{d \Big[ \left(\nicefrac{\Delta \rho}{\rho}\right)_{_{xy}}(H_z) - \left(\nicefrac{\Delta \rho}{\rho}\right)_{_{yx}}(H_z) \Big]}{d \Big[ \epsilon_{zz} \Big]} \bigg|_{\epsilon_{_{xx}} + \epsilon_{_{yy}}=0} \\
& \quad = 2 m_{xy,zz}
\end{split} \\
\begin{split}
&{} \Lim{(\epsilon_{xz} + \epsilon_{zx}) \to 0} \frac{d \Big[ \left(\nicefrac{\Delta \rho}{\rho}\right)_{_{yz}}(H_z) - \left(\nicefrac{\Delta \rho}{\rho}\right)_{_{zy}}(H_z) \Big]}{d \Big[ \epsilon_{xz} + \epsilon_{zx} \Big]} = 2 m_{yz,zx}
\end{split} \\
\begin{split}
&{} \Lim{(\epsilon_{yz} + \epsilon_{zy}) \to 0} \frac{d \Big[ \left(\nicefrac{\Delta \rho}{\rho}\right)_{_{zx}}(H_z) - \left(\nicefrac{\Delta \rho}{\rho}\right)_{_{xz}}(H_z) \Big]}{d \Big[ \epsilon_{yz} + \epsilon_{zy} \Big]} = 2 m_{yz,zx}
\end{split}
\end{align}
\end{subequations}

\noindent As mentioned above, since quantities with the same symmetry can mix into each other and there are two forms of $\epsilon_{_{A_{1g}}}$, the ratios in \eqref{eq:18} involving $\epsilon_{xx}+\epsilon_{yy}$ and $\epsilon_{zz}$ are only equal to the indicated coefficients provided that the other $A_{1g}$ strain is constrained to vanish, which we have denoted by $\big|_{\epsilon_{_{A_{1g}}}=0}$. Achieving such constraints in practice might be challenging, but we emphasize that in this discussion we are more concerned with formal definitions than with the practical means that might be employed to realize such an experiment.

\section{Conclusion}

The primary goal of this work has been twofold.  First, we have tried to explain the general ways in which the elastoresistivity tensor is constrained due to the structure of the resistivity and strain tensors and also the point group symmetry of the crystal.  We focused in detail on the tetragonal point group $D_{4h}$, but extensions to other point groups would proceed in an analogous way.  Our treatment also readily incorporates the presence of a magnetic field.  Second, our motivation in pursuing elastoresistance measurements has been to elucidate the role of electronic nematicity (broken rotational symmetry driven by electronic correlations) in a number of strongly correlated electron systems.  To this end, we have discussed how elastoresistivity coefficients can be connected to thermodynamic susceptibilities in order to characterize the symmetry of an order parameter for a continuous phase transition.

\begin{acknowledgments}

The authors thank S. A. Kivelson and S. Raghu for helpful discussions.  Work at Stanford was supported by the U.S. DOE, Office of Basic Energy Sciences, under contract DEAC02-76SF00515.

\end{acknowledgments}


\begin{thebibliography}{99}
\bibitem{footnote2} Strictly, the Onsager relation applies to $\rho_{ij}(\boldsymbol H)$; however, it carries over to $\left( \nicefrac{\Delta \rho}{\rho} \right)_{ij}(\boldsymbol H)$ as well.  Since Onsager obtains for both $\rho_{ij}(\hat{\epsilon} = \hat{0})$ and $\rho_{ij}(\hat{\epsilon})$, it necessarily holds for $\Delta \rho_{ij}(\hat{\epsilon}) \equiv \rho_{ij}(\hat{\epsilon}) - \rho_{ij}(\hat{\epsilon} = \hat{0})$ too.  Using the normalization scheme of Eq. \eqref{eq:4}, it follows that

\begin{align}
\label{eq:19}
& \left( \nicefrac{\Delta \rho}{\rho} \right)_{ij}(\boldsymbol H) = \left( \nicefrac{\Delta \rho_{ij}(\boldsymbol H)}{\sqrt{\rho_{ii}(\boldsymbol H)}\sqrt{\rho_{jj}(\boldsymbol H)}} \right) \\
& \quad = \left( \nicefrac{\Delta \rho_{ji}(-\boldsymbol H)}{\sqrt{\rho_{ii}(-\boldsymbol H)}\sqrt{\rho_{jj}(-\boldsymbol H)}} \right) = \left( \nicefrac{\Delta \rho}{\rho} \right)_{ji}(-\boldsymbol H). \nonumber
\end{align}
\bibitem{footnote3} We note that in our compactified notation where the fourth rank tensor $m_{ij,kl}$ is represented as a single matrix, Eq. \eqref{eq:5} is simply a basis transformation of matrices $m_{\alpha\beta} = \Lambda_{\alpha\mu}\Lambda_{\beta\nu}m_{\mu \nu}$, where $\Lambda_{\alpha\mu}$ is a Kronecker product of the coordinate transformation matrices $O$ by $\Lambda = O \otimes O$. In particular, in the basis of strain where $\vec{\epsilon} = (\epsilon_{xx},\epsilon_{yy},\epsilon_{zz},\epsilon_{yz},\epsilon_{zy},\epsilon_{zx},\epsilon_{xz},\epsilon_{xy},\epsilon_{yx})^{T}$ (and likewise for change in resistivity), the transformation $\Lambda$ is a $9 \times 9$ matrix given by
\unexpanded{\begin{equation}\label{eq:20}
\Lambda = \begin{pmatrix}
O_{xx} O_{xx}& O_{xx} O_{yy} & \ldots  & O_{xx} O_{yx}\\
O_{yy} O_{xx} & O_{yy} O_{yy} &\ldots & O_{yy} O_{yx}\\
\vdots &\vdots &\ddots&\vdots\\
O_{yx}O_{xx} & O_{yx}O_{yy}&\ldots &O_{yx}O_{yx}
\end{pmatrix}.
\end{equation}}
The specific values of the elements of the $O$ matrices depend on how the particular transformation operates on the basis vectors.  As an illustrative example, for the $\sigma_z$ transformation, $O_{xx} = O_{yy} = 1$, $O_{zz} = -1$, and all other $O_{ij} = 0$.
\bibitem{footnote5} Improper rotations are compositions of reflections and rotations, and so they do not impose additional constraints on the elastoresistivity tensor beyond those from the reflections and rotations individually.
\bibitem{footnote7} If $\Gamma_{i}$ transformed like the identity of the point group, no symmetry would be broken, and hence the order parameter would not be well-defined.
\bibitem{footnote8} In enumerating the irreps of $D_{4h}$ with representations in strain, we have chosen $\frac{1}{2}(x^2+y^2)$ and $z^2$ as the two quadratic basis functions for $A_{1g}$.  This is a standard choice, but there is another equally standard choice involving the basis functions $\frac{1}{3}(x^2+y^2+z^2)$ and $\frac{1}{6}(2z^2-x^2-y^2)$, which when represented in strain corresponds to decomposing an arbitrary deformation into a volume-changing component and a volume-preserving component.
\end{thebibliography}
\end{document}